\documentclass[12pt]{article}

\usepackage{amsmath}
\usepackage{cite}

\newcommand{\mys}[1]{\section{#1}
	\setcounter{equation}{0}}

\newlength{\dummysp}
\newcommand{\tr}{\mathop{{\hbox{Tr} \, }}\nolimits}

\newcommand{\half}{{\frac{1}{2}}}

\newcommand{\beq}{\begin{eqnarray}}
\newcommand{\eeq}{\end{eqnarray}}
\newcommand{\nnn}{ \nonumber \\ }

\newcommand{\Ucal}{{\cal U}}

\newcommand{\Acal}{{\cal A}}
\newcommand{\Ccal}{{\cal C}}

\newcommand{\chib}{{\bar \chi}}

\newcommand{\ord}[1]{{{\cal O}(#1)}}
\newcommand{\gappeq}{\mathrel{\rlap {\raise.5ex\hbox{$>$}}
{\lower.5ex\hbox{$\sim$}}}}
\newcommand{\lappeq}{\mathrel{\rlap{\raise.5ex\hbox{$<$}}
{\lower.5ex\hbox{$\sim$}}}}
\newcommand{\myref}[1]{(\ref{#1})}

\newcommand{\ben}{\begin{enumerate}}
\newcommand{\een}{\end{enumerate}}

\newcommand{\fourth}{\frac{1}{4}}

\newcommand{\bit}{\begin{itemize}}
\newcommand{\eit}{\end{itemize}}

\newcommand{\obf}{{\bf 1}}

\newcommand{\muhat}{{\hat \mu}}
\newcommand{\qbar}{{\bar q}}

\def\[{\left [}
\def\]{\right ]}
\def\({\left (}
\def\){\right )}

\begin{document}
\begin{titlepage}

\renewcommand{\thefootnote}{\fnsymbol{footnote}}

\hfill Dec.~30, 2005

\hfill hep-lat/0507002

\vspace{0.45in}

\begin{center}
{\bf \Large Toward a systematic analysis \\ \vskip 10pt
of the fourth-root trick}
\end{center}

\vspace{0.15in}

\begin{center}
{\bf \large Joel Giedt\footnote{{\tt Presently: giedt@physics.umn.edu}}}
\end{center}

\vspace{0.15in}

\begin{center}
{\it University of Toronto, Department of Physics \\
60 St. George St., Toronto ON M5S 1A7 Canada}
\end{center}

\vspace{0.15in}

\begin{abstract}
In this note I briefly discuss ideas
related to the so-called fourth-root trick.
A decomposition of the ``rooted'' fermion
effective action into Wilson fermions and a nonlocal,
lattice spacing suppressed functional is presented,
complete with link interactions.  Some
proposals are given for analytical, nonperturbative
studies of the fourth-root trick.
\end{abstract}

\end{titlepage}

\renewcommand{\thefootnote}{\arabic{footnote}}
\setcounter{footnote}{0}

\mys{Introduction}

Lattice QCD with improved staggered fermions (SF)
has recently enjoyed publicity
for its ability to correctly reproduce many aspects
of hadron physics with reasonable 
accuracy \cite{Davies:2003ik,ptart}.  However, criticism
has been leveled at the approach 
\cite{Jansen:2003nt,DeGrand:2003xu,Neuberger:2004ft}
due to, among other things, 
the use of the {\it fourth-root trick} (FRT).  
In this note, I briefly review aspects of this
issue, and mention some ideas in relation to it.

The FRT is used because staggered fermions, or, Kogut-Susskind fermions 
\cite{Kogut:1974ag,Susskind:1976jm,Sharatchandra:1981si}, 
do not entirely overcome
the fermion doubling problem.  Rather, they reduce the
number of continuum modes from 16 to 4.  
These 4 modes are referred to as {\it tastes,}
so as to distinguish them from the $N_f$ flavors
in the continuum (target) theory.  To
estimate the fermion measure of $N_f$
continuum flavors, one takes the power $N_f/4$ of the
fermion determinant in the definition of
the functional integral.  This is
what I will refer to as the fourth-root trick,
although it is often called the square-root trick
since in QCD two light flavors are used.

In the case of free staggered fermions,
the eigenvalues $\omega_k$ of the fermion matrix $M_{SF}$
are 4-fold degenerate, corresponding to
symmetries that relate the 4 tastes.  Thus
there appears to be nothing wrong with the FRT:
\beq
\det M_{SF}^{1/4} = \(\prod_k \omega_k^4 \)^{1/4} = \prod_k \omega_k
\equiv \det M_{eff}.
\eeq
That is, one effectively weights by the eigenvalue
spectrum of some other ``fermion matrix'' $M_{eff}$.
It has been shown by Shamir \cite{Shamir:2004zc}
that RG blocking allows one to write (as suggested in
\cite{Adams:2004mf})
\beq
\det M_{eff} = \det D_{RG} \det T^{1/4} .
\eeq
Here, $D_{RG}$ is a local lattice Dirac operator with a single
flavor.  $T$ is a local operator that only contains UV
modes.  Thus nonlocality contained in $\det T^{1/4}$ factorizes into
a harmless overall factor once the UV modes are 
integrated out.  The long-distance effective
action is local.

The problem is that one is not interested in
free staggered fermions.  Rather, in QCD the
gauge-covariant staggered fermion operator
contains interactions through the link fields $U_\mu(r)$.
For a generic configuration of the link fields,
the 4-fold degeneracy of the eigenvalues of
the fermion matrix is destroyed:
\beq
\prod_k \omega_k^4 \to \prod_k \prod_{\alpha=1}^4
\omega_{k,\alpha}(U) .
\eeq
In this circumstance, one has a right to
worry about the FRT.

The splitting of eigenvalues occurs
due to {\it taste violating} (TV) interactions.  
In an expansion of link bosons about the
identity, these arise from higher lattice-derivative, 
irrelevant operators.
That is, the effect is suppressed by a power of
the lattice spacing $a$ that depends on the level
of improvement of the lattice action,
and increases with the amplitude of UV components
of the link configuration.
The asymmetric mixing of tastes that occurs in the presence of
a nontrivial link configuration lifts the 4-fold
degeneracy.  Numerical studies of these splittings
have been performed for dynamically generated link configurations.
For example, in \cite{Durr:2004as} it was seen that
smearing of links---which reduces the TV effects
by smoothing out UV link fluctuations---can
render the IR eigenvalues ``near degenerate ({\it sic}).''
TV effects are regarded as a
source of systematic error that can be
reduced with enough effort.  Since it has
been observed that many physical observables
are insensitive to a truncation of
UV eigenvalues of the lattice Dirac operator (e.g.~\cite{Duncan:1998gq}),
the nonlocalities associated with ``rooting'' 
a spectrum with nearly degenerate IR eigenvalues and
nondegenerate UV eigenvalues are perhaps harmless.
While this sort of argument is somewhat reassuring,
clearly it would be preferable
to rigorously address the consistency
of the ``rooted'' theory.

Improved staggered fermions are used because they are efficient,
cost effective, and seem to give good results.  This
raises simple questions that we would like to
answer:  Why does the FRT work in practice?  At what point
might it fail?  For instance, is there a limit
on its applicability due to a breakdown
of unitarity at some short distance scale?

I now summarize the remainder of this note:
\bit
\item
In \S\ref{s:out}, I outline various perspectives
on the FRT.  
Firstly, in \S\ref{pvw} I remark on the effects of the FRT
in perturbation theory.  I reach the conclusion
which has been stated many times:  the FRT poses
no problem for weak coupling perturbation theory
about the $A_\mu=0$ vacuum.  I present
details as to why this is a reasonable conclusion.
Secondly, in \S\ref{npi} I
speculate as to analytical, nonperturbative
studies that could be done to study the FRT.
Thirdly, in \S\ref{wfd} I present an unusual decomposition
of the ``rooted'' fermion effective action,
in terms of Wilson fermions, and a nonlocal
TV operator that has explicit suppression
by the lattice spacing.
Applications of this decomposition, which isolates the taste-violating
part of the theory, are left to future work.
\item
In \S\ref{s:det}, I discuss some details 
of the staggered fermion action.  
A translation into the taste basis, without
recourse to expansion about $U_\mu=1$, is described.
The explicit form of the decomposition into
Wilson fermions and a TV correction is given
in the taste basis.
\item
In \S\ref{s:con}, I conclude with a summary.
Directions for future research are recapitulated.
\eit

\mys{Perspectives}
\label{s:out}

\subsection{The perturbative view}
\label{pvw}
Let me begin with conventional
perturbation theory.  One expands about the
$U_\mu(r)=\exp(iagA_\mu(r)) \to 1$ limit.  First I write
\beq
M_{SF}(U) = M_{SF}(1) + [ M_{SF}(U) - M_{SF}(1) ]
\equiv M_{SF}(1) + \Delta M_{SF}(U) .
\label{cand}
\eeq
Note that $\Delta M_{SF}(U) = \ord{igA_\mu}$
and contains, among other things, the minimal fermion-boson vertex.
With the FRT applied, the fermion measure is represented by
the following effective contribution to the action:
\beq
S_{eff} & = &
- \frac{N_f}{4} \tr \ln \[ M_{SF}(1) + \Delta M_{SF}(U) \] \nnn
&=& -\frac{N_f}{4} \tr \ln \[ 1 + M_{SF}^{-1}(1) \Delta M_{SF}(U) \]
+ {\rm const.}
\label{cexp}
\eeq
One expands $\exp (-S_{eff})$ in $\Delta M_{SF}(U)$
to obtain a series of terms, each $\tr$ multiplied
by $N_f/4$.  One further expands $\Delta M_{SF}(U)$
in terms of $U_\mu=\exp i ag A_\mu$ to obtain
weak coupling perturbation theory and the continuum
limit.  Since each $\tr$ corresponds to a
fermion loop, the effect of the FRT is seen to
be as follows:  each diagram of ordinary SF perturbation
theory is multiplied by $(N_f/4)^{n}$, with $n$ the
number of fermion loops.  

If perturbation theory
is consistent at $N_f=0$ mod $4$ (i.e., for ordinary
SF perturbation theory), then it is difficult to
see how any problems would arise for values of
$N_f$ in-between.  
E.g., let $Z_i,m,g,\ldots$ be chosen to renormalize
the theory for $N_f=0$ mod $4$.  These then become
functions of $N_f/4$, with a natural extension to
arbitrary $N_f$.  It is conceivable that 
cancellations between divergent diagrams happen at $N_f=0$ mod $4$
but not at other values.  That would indicate
the need for additional counterterms; one
might worry whether or not they are always local,
given the apparent nonlocal nature of the
rooted SF matrix.  However,
the symmetries of the SF action are still operative
with the FRT applied, so the form of counterterms
is similarly restricted.  Also, it can be
seen that the perturbation series is constructed
entirely from local vertex operators, since
the only change is a factor of $N_f/4$ for
each fermion loop.  This is enough to exclude
the possibility of nonlocal counterterms.

Finally, renormalization of composite operators could
be differerent when $N_f \not= 0$ mod $4$,
due to a lack of certain cancellations.  But, I have
no example to cite.  

The conclusion I draw from these considerations
is in agreement with ``standard lore'':  the
FRT poses no problem in perturbation theory.

\subsection{Nonperturbative ideas}
\label{npi}
Perhaps nonperturbative effects of the FRT might
be accessible through an instanton calculus, or
some other semiclassical expansion.  Whereas
it has been argued above that the FRT poses
no problems for perturbation theory about
the $A_\mu=0$ vacuum, perturbation theory
about some other background could conceivably
yield different conclusions.  The point
is to compare to results obtained from other approaches
in a similar regime:  from the continuum,
Ginsparg-Wilson fermions, etc.  Finding any
discrepancy and understanding its origin
would teach us valuable things.  Finite
system volume, perhaps in conjunction with
twisted boundary conditions, could help to
exert greater theoretical control over such calculations.

It might also be of interest to look at strong
coupling expansions with the FRT applied.  Some
nonperturbative features, such as chiral symmetry
breaking, can be studied by this approach; e.g.~\cite{vandenDoel:1983mf}.
One could compare expectations for the spectrum of mesons and
baryons for a theory with $N_f$ flavors to
what occurs in the strong coupling limit
with fermion measure \myref{cexp}.

\subsection{The Wilson fermion decomposition}
\label{wfd}
In this decomposition, I organize the effective fermion
measure in such a way that TV effects are isolated
for systematic study, and suppression by the
lattice spacing is explicit.

Let $M_{WF}$ be a single flavor Wilson fermion (WF) matrix.
Then I decompose the staggered matrix as follows:
\beq
M_{SF} = M_{WF} \otimes \obf_4 + a M_{TV} ,
\eeq
where $M_{TV}$ contains all the TV
effects.  This decomposition is possible
because in the taste (flavor) basis
for staggered fermions 
\cite{Gliozzi:1982ib,Kluberg-Stern:1983dg}, 
the terms that are not
suppressed by $a$ are just four identical
flavors of the naive lattice Dirac
matrix. (With gauge interactions included,
some reinterpretation of link variables
is required; c.f.~\myref{rflv}.  This will be explained in detail
below.) Once this decomposition has been effected,
I note that
\beq
\det M_{SF}^{N_f/4} = \det M_{WF}^{N_f} \exp \frac{N_f}{4} 
\tr \ln \[ \obf + a (M_{WF}^{-1} \otimes \obf_4) M_{TV} \] .
\eeq
It can be seen that this arrangement sequesters all the TV effects and
nonlocality into $\ord{a}$ correction terms.\footnote{
I note that the operator 
$[(M_{WF}^{-1} \otimes \obf_4) M_{TV}]^{1/4}$
is the same one that Adams has studied in \cite{Adams:2004mf}.
In this respect, his proposal is related to the one
presented here.}  The fermions
of the theory now consist of $N_f$ degenerate flavors, each with
a conventional matrix $M_{WF}$.  The correction term
is a nonlocal functional of the the link fields.

One disadvantage of the Wilson fermion decomposition is that
it will typically obscure the
staggered fermion symmetries, such as the
one that prevents additive mass renormalization.
These are hidden in the induced tranformations
of the correction term.

Finally, I remark that the Wilson fermion
is not the unique choice for the type of decomposition
described here.  Any other fermion that consists of
the naive fermion plus $\ord{a}$ terms would
also work.  It remains to be shown that this
decomposition has a useful application.  I
will not do that here, but hope to find one
in the future.

\mys{Taste basis details}
\label{s:det}

The fermion action I start with is the single-flavor
staggered fermion, written in the conventions of 
Kluberg-Stern et al.~\cite{Kluberg-Stern:1983dg}.
It is just (with obvious notations and summation conventions)
\beq
S_{SF} &=& -\half a^3 \alpha_\mu(r)
\[ \chib(r) U_\mu(r) \chi(r + \muhat) 
+ \chib(r + \muhat) U_\mu^\dagger(r) \chi(r) \] \nnn
&& + \; i m a^4 \chib(r) \chi(r).
\label{sfacti}
\eeq
The phases are, as usual,
$\alpha_\mu(r) = (-1)^{\sum_{\nu < \mu} r_\nu}$.
One associates a $(2a)^4$ hypercube
with each site $r=2y$, where $y$ has integral entries.  Sites
contained in the hypercube are labeled on the original
lattice by
\beq
r=2y+\eta, \quad \eta \in K \equiv \{ (0^4), (\underline{1,0^3}),
(\underline{1^2,0^2}), (\underline{1^3,0}), (1^4) \}.
\eeq
Powers indicate how many times a 0 or 1 appears
and underlining indicates that
all permutations of entries are to be included.

To proceed further, one defines fermion fields
for each point in the hypercube and transforms
to the (covariant, position space) taste basis: $\chi(r),\chib(r) \to
q^{\alpha a}(y),\qbar^{a \alpha}(y)$.  Since this is all very well-known,
I just summarize the ingredients:
\beq
&& \chi(2y+\eta) = (-1)^{\sum_\mu y_\mu} \chi_\eta(y), \quad
\chib(2y+\eta) = (-1)^{\sum_\mu y_\mu} \chib_\eta(y),
\nnn
&& \Gamma_\eta = \gamma_1^{\eta_1} \gamma_2^{\eta_2}
\gamma_3^{\eta_3} \gamma_4^{\eta_4}, \quad
\{ \gamma_\mu, \gamma_\nu \} = -2\delta_{\mu\nu}, 
\nnn
&& U_{\mu,\eta}(y) = U_\mu(2y+\eta), \quad
\Gamma_{\eta, \eta'}^\mu = \fourth \tr (\Gamma^\dagger_\eta
\gamma_\mu \Gamma_{\eta'}), 
\nnn
&& U_\eta(y) \equiv  U_1^{\eta_1}(2y) U_2^{\eta_2}(2y+\eta_1)
U_3^{\eta_3}(2y+\eta_1+\eta_2) U_4^{\eta_4}(2y+\eta_1+\eta_2+\eta_3) ,
\nnn
&& \chi_\eta(y) = 2 U_\eta^\dagger(y) \tr \( \Gamma_\eta^{\dagger} q(y) \),
\quad \bar \chi_\eta(y) = 2 \tr \( \Gamma_\eta \qbar(y) \) U_\eta(y) .
\eeq

Using various well-known identities, I have found that
\myref{sfacti} becomes:
\beq
S_{SF} &=& 2a^3 \left\{ 
\qbar(y) C_\mu(y) q(y+\muhat)
+ \qbar(y+\muhat) C_\mu^{\dagger}(y) q(y) 
\right. \nnn
&& \left. + \; 
\qbar(y) \( A(y) + A^{\dagger}(y) \) q(y) 
+ 8iam \qbar(y) (\obf \otimes \obf) q(y) \right\} .
\label{lsfa}
\eeq
I have defined the following link-dependent structures (in
these two equations all sums are explicit):
\beq
A^{\alpha \beta; a b}(y) &=& 
\sum_\mu \sum_{\eta,\eta' \in K}
\delta_{\eta+\muhat,\eta'}
\Gamma_\eta^{\alpha a} \Gamma_{\eta'}^{\dagger b \beta}
\Gamma_{\eta, \eta'}^\mu
U_\eta(y) U_{\mu,\eta}(y) U_{\eta+\muhat}^\dagger(y) ,
\nnn
C_\mu^{\alpha \beta; a b}(y) &=& 
\sum_{\eta,\eta' \in K}
\delta_{\eta-\muhat,\eta'}
\Gamma_\eta^{\alpha a} \Gamma_{\eta'}^{\dagger b \beta}
\Gamma_{\eta, \eta'}^\mu
U_\eta(y) U_{\mu,\eta}(y) U_{\eta-\muhat}^\dagger(y+\muhat) .
\eeq
Greek and latin indices correspond to spinor and taste labels
respectively.  
These have been suppressed in \myref{lsfa},
but are easy to put back in.
For instance, $\qbar(y) A(y) q(y) =
\qbar^{a \alpha}(y) A^{\alpha \beta; a b}(y) q^{\beta b}(y)$.

It is not difficult to show that (on the original
lattice) $A(y)$ is a parallel
transporter from $2y$ back to $2y$ along a sum
of paths.  These paths traverse the $(2a)^4$ hypercube
that is associated with the site $y$ of the doubled lattice.
$A(y)$ thus transforms as a site variable at $y$
on the doubled lattice.  $C_\mu(y)$ transports
from $2y$ to $2(y+\muhat)$ along a sum of paths
in the hypercube.  It is thus a link variable
from $y$ to $y + \muhat$ on the doubled lattice.
This is consistent with the transformation
properties of the quark fields appearing
in~\myref{lsfa}.

Consider the action obtained by expanding in $ag$
the links $U_\mu(r)=\exp[iagA_\mu(r)]$
that appear implicitly through $A(y)$ and $C_\mu(y)$
in \myref{lsfa}.
Kluberg-Stern et al.~give this continuum approximation
to the interacting theory up to
$\ord{a^2}$ corrections \cite{Kluberg-Stern:1983dg}.  This exposes
the leading irrelevant operators, which
are suppressed by a single power of $a$.
It is a simple matter 
to decompose their expression into 4 degenerate
tastes of WFs (lattice spacing $2a$) and TV terms:
\beq
S_{SF} &=& S_{WF(4)} + a S_{TV}  \nnn
&\equiv& (2a)^4 \qbar(y) \[ M_{WF(4)} + a M_{TV} \] q(y) , \nnn
M_{WF} &=& \gamma_\mu D_\mu -i a D_\mu D_\mu + im, \quad
M_{WF(4)} = M_{WF} \otimes \obf_4 , \nnn
M_{TV} &=& \sum_\mu \[ \gamma_5^\dagger 
\otimes (t_\mu^\dagger t_5^\dagger) +i \obf_4 \otimes
\obf_4 \] D_\mu^2 
- \sum_{\mu \nu} \frac{ig}{4} T_{\mu \nu} F_{\mu \nu}
+ \ord{a}, \nnn
T_{\mu \nu} &=& (\gamma_\mu - \gamma_\nu) \otimes \obf_4
+ \half \gamma_5^\dagger [\gamma_\mu , \gamma_\nu]
\otimes ((t_\mu + t_\nu)^\dagger t_5^\dagger) .
\label{wftvc}
\eeq
Here, $D_\mu$ is the gauge-covariant derivative
and $F_{\mu \nu}$ is the field-strength.

In the same spirit,
I can make Wilson fermions manifest
in \myref{lsfa}, working entirely in terms
of link variables.
I define links that connect sites of the doubled lattice:
\beq
\Ucal_\mu(y) \equiv U_\mu(2y) U_\mu(2y+\muhat) .
\label{rflv}
\eeq
This permits one to define the action of 4 flavors of WF
on the doubled lattice:
\beq
&& S_{WF(4)} = 4a^3 \left\{ \qbar(y)  \( (i+\gamma_\mu) \otimes
\obf \) \Ucal_\mu(y) q(y+\muhat) \right. \nnn
&& \quad \left. + \; \qbar(y+\muhat)  \( (i-\gamma_\mu) \otimes
\obf \) \Ucal_\mu^\dagger(y) q(y)
+i(4ma-8) \qbar(y) ( \obf \otimes \obf ) q(y) \right\} .
\eeq

If I add and subtract this from \myref{lsfa}, I then
have $S_{SF}=S_{WF(4)}+aS_{TV}$, where
\beq
S_{TV} &=& (2a)^4 \left\{ \qbar(y) \Ccal_\mu(y) q(y+\muhat)
+ \qbar(y+\muhat) \Ccal_\mu^\dagger(y) q(y) 
+ \qbar(y) \Acal(y) q(y) \right\}, \nnn
\Acal(y) &\equiv& \frac{1}{8a^2}
\[ A(y) + A^\dagger(y) \] 
+ \frac{2i}{a^2} (\obf \otimes \obf),
\nnn
\Ccal_\mu(y) &\equiv&
\frac{1}{8a^2} C_\mu(y)
- \frac{1}{4a^2} \( (i+\gamma_\mu) \otimes
\obf \) \Ucal_\mu(y) .
\label{lsfb}
\eeq

\mys{Discussion}
\label{s:con}

I have given an argument in support of the
conclusion that the FRT poses no problem in
perturbation theory about the $A_\mu=0$ vacuum.
The consistency of the FRT needs to be
studied further at a nonperturbative level.

I have suggested a semiclassical analysis about nontrivial link
configurations; for example, ones that have nonzero 
topological charge.
Perhaps that might allow for an examination of nonperturbative
TV vis-\'a-vis the fourth-root trick.  The point
is to compare to results obtained from other approaches
in a similar regime.  It would also be interesting
to examine the strong coupling limit.  Studies in
these directions are currently in progress.

It is not clear to me whether or not the
questions of global topology suggested in
\cite{Neuberger:2004ft} can be addressed by semiclassical or
strong coupling methods.  However, many other important questions
of consistency may be accessible through
the techniques envisaged here.

Finally, I hope to present some useful applications
of the Wilson fermion decomposition at a later date.

\vspace{15pt}

\noindent
{\bf \Large Acknowledgements}

\vspace{5pt}

\noindent
I thank Erich Poppitz, George Fleming,
Rajan Gupta and Gunnar Bali for helpful discussions.
This work was supported by the National Science and Engineering 
Research Council of Canada and the Ontario 
Premier's Research Excellence Award.

\end{document}